\documentclass[conference]{IEEEtran}
\usepackage{graphicx}
\usepackage{multirow}
\usepackage{amssymb}
\usepackage{epstopdf}
\usepackage{lipsum}
\usepackage{stfloats}
\newcommand\Mark[1]{\textsuperscript#1}

\ifCLASSINFOpdf
\else
\fi
\begin{document}

\title{A Flow Sensitive Security Model \\for \\Cloud Computing Systems}

\author{
    \IEEEauthorblockN{Wen Zeng\Mark{1}, Chunyan Mu\Mark{2}, Maciej Koutny\Mark{1}, Paul Watson\Mark{1}}
    \IEEEauthorblockA{\Mark{1}School of Computing Science, Newcastle University \\
    Newcastle upon Tyne NE1 7RU, U.K.
    \\wen.zeng.wz@gmail.com, \{maciej.koutny, paul.watson\}@ncl.ac.uk}
    \IEEEauthorblockA{\Mark{2}LIAFA, Universit\'{e} Paris Diderot, Paris 7, France
    \\chunyan.mu@liafa.univ-paris-diderot.fr}
}

\maketitle
\IEEEpeerreviewmaketitle

\section{Background and Motivation}

The extent and importance of cloud computing is rapidly increasing due to the ever increasing
demand for internet services and communications.
Instead of building individual information technology infrastructure to host databases or software,
a third party can host them in its large server clouds.  Large organizations may
wish to keep sensitive information on their more restricted servers rather than in the public cloud.
This has led to the introduction of federated cloud computing (FCC)  in which
both public and private cloud computing resources
are used~\cite{Watson11}.

A federated cloud is the deployment and management of
multiple cloud computing services with the aim of  matching business needs. Large number of data and services are required to be allocated in different clouds for business concerns, which creates security risks, due to the operational independence of clouds and their geographic distribution. As a result, it is very hard for an
organization to track and control the information flow in the system.
It is therefore necessary to develop a formal
model describing the information flow security within an federated cloud system (FCS),
making the information and data traceable.

There exist different methods for addressing workflow security;
for example, \cite{Knorr01} applied the Bell-LaPadula model to 
address this problem. However,
the deployment of blocks within a workflow across a set of computational resources 
has not been considered.
The paper~\cite{Watson11} proposed to partition workflows
over a set of available clouds in such a way that security
requirements are met. However,
the concurrency of the events or the execution of tasks in the system was not 
considered. Therefore, the goal of this paper~\cite{Zeng13} is to analyze the 
security of information flow in FCSs.

\section{Contribution}

The main contributions of this paper are summarized below: 

Firstly, security lattices is introduced for the components 
of a cloud system as well as for sets of individual clouds.

Secondly, a flow sensitive security model (FSSM) is 
introduced to capture the information flow in 
a federated cloud computing systems. The state transitions of 
the model can  be analyzed to verify that they satisfy 
conditions of a given security policy such as non-interference 
properties, Bell-Lapadula rules for confidentiality considerations, 
and user-specified policies. This model can be captured by Coloured Petri Nets (CPNs).

Finally, the opacity of FSSMs is investigated. 
Opacity is a uniform approach for
describing security properties expressed as predicates~\cite{Bryans2008}. 
A predicate is opaque if an observer of the system is unable to
determine the truth of the predicate in a given run of the system. 
In  service-oriented distributed computing systems, 
information sharing means that the behaviour of one cloud user 
may appear visible to other cloud users or adversaries, and
observations of such behaviours can potentially help adversaries
to build covert channels. Therefore, opacity is an
effective way  to 
measure the amount of information related to a service 
that might be exposed to other users or adversaries.

\section{Conclusions}\label{sec:conclusion}

A flow sensitive security model is presented 
to analyse information
flow in federated cloud systems. Each cloud and the 
entities of the cloud system are classified into
different security levels which form a security lattice. 
Opacity --- a general technique for unifying
security properties --- turns out to be a promising 
analytical technique in the context of cloud computing systems.
The proposed approach can help to track and control 
the secure information flow in federated
cloud systems. It  can also be used to analyze
the impact of different resources allocation strategies.



\bibliographystyle{IEEEtran}
\bibliography{IEEEabrv,paper}

\end{document}